\documentclass[%
preprint,
longbibliography,
amsmath,amssymb,
aps,
prd,
sort&compress,
merge
]{revtex4-2}

\usepackage{graphicx}
\usepackage{bm}
\usepackage{booktabs}
\usepackage{xcolor}
\usepackage{hyperref}

\begin{document}

\title{State-Dependent Visibility of Non-Commutative Ordering in Quantum Dynamics}

\author{Ji-Chong Yang}
\email{yangjichong@lnnu.edu.cn}
\thanks{Corresponding author}
\author{Gui-Qi Hu}

\affiliation{Department of Physics, Liaoning Normal University, Dalian 116029, China}
\affiliation{Center for Theoretical and Experimental High Energy Physics, Liaoning Normal University, Dalian 116029, China}

\date{\today}

\date{\today}

\begin{abstract}
A nonzero commutator proves that two orderings differ as operators, but it does not ensure that a physical state can reveal the difference.  
We ask when non-Abelian ordering information becomes dynamically invisible.  
For Hermitian operators $B$ and $C$, we compare the evolutions generated by the opposite-order products $M=(B+iC)(B-iC)$ and $\widetilde M=(B-iC)(B+iC)$, and define their operational visibility from the minimum overlap of the output states over a normalized time window.  
This visibility bounds the difference produced by the two orderings in every observable on the chosen state.  
An exact one-qubit solution shows that the same fixed pair can be perfectly invisible in one state and visible in another.
We then keep the ordered generators fixed and vary only the many-body ground state across a quantum phase transition.  
The same ordering difference is nearly invisible in one regime and clearly visible in the other.  
Moreover, states with identical leading quadratic decay can develop sharply different finite-time visibility because their first distinction appears at higher order.  
The effect persists across distinct operator pairs and coefficient perturbations.  
Thus dynamical Abelianization is a property of the state-dependent process, i.e., non-Abelian ordering information can become operationally inaccessible even though the underlying operators remain non-commuting.
\end{abstract}

\maketitle

\section{Introduction}
\label{sec:intro}

In studies of non-Abelian gauge theories, ``Abelian dominance'' refers to a striking observation.  
After an Abelian projection, the retained Abelian field can reproduce several long-distance quantities of the full theory, including much of the string tension~\cite{Suzuki1990,Shiba1994,Stack1994,Bali1996}.  
These studies are closely connected with monopole dominance.  
Monopole condensation is central to the dual-superconductor picture of confinement~\cite{tHooft1981,Kronfeld1987,DiGiacomo2000,DAlessandro2007}, whereas the infrared suppression or effective mass of off-diagonal gluons is more directly connected with why the Abelian field can dominate long-distance observables~\cite{Kondo1998,Amemiya1999}.  
This raises another question.  
Must the ordering carried by a non-Abelian operation remain distinguishable in the dynamics, or can two operations that differ in ordering become indistinguishable on a given input state?
Abelian dominance motivates our study, but our primary question is whether the ordering information that distinguishes Abelian from non-Abelian multiplication can become dynamically invisible in a physical state.  
We call this state-dependent suppression of ordering information dynamical Abelianization.  

Where does the distinction between Abelian and non-Abelian multiplication enter the dynamics?
A complete Hamiltonian specifies the entire dynamics.
Equivalently, a pure lattice gauge theory can be specified by a gauge action built from plaquettes, each of which contains an ordered product of link variables.  
For an Abelian gauge group, the link variables commute and the product carries no ordering information.  
For a non-Abelian gauge group, the order cannot be removed.  
Thus, in pure lattice gauge dynamics, the Abelian--non-Abelian distinction is whether the multiplication order of the link variables matters.  
This ordering dependence appears in the Hamiltonian generators that determine time evolution.  
It is therefore the generated operation, rather than only an operator-level number, that we focus on in this work.  
For two non-commuting operators $A$ and $B$, the expectation value $\langle[A,B]\rangle$ can diagnose state-dependent commutator information, but it does not by itself answer whether two ordered dynamical processes produce distinguishable output states.  
We therefore compare the operations generated by two different orderings and ask whether their outputs can be distinguished on a given input state.  
Their state- and time-dependent distinguishability is what we call the \emph{operational visibility of non-commutativity}.

To turn this idea into a concrete ordering test, we need two Hermitian generators that contain the same factors in opposite orders.
Take two Hermitian operators $B$ and $C$, and define $A_1=B+iC$ and $A_2=B-iC=A_1^\dagger$.  
Then $M=A_1A_2$ and $\widetilde M=A_2A_1$ are both Hermitian, while $M\neq\widetilde M$ whenever $[B,C]\neq0$.  
They generate two unitary operations, $U(t)=e^{-itM}$ and $\widetilde U(t)=e^{-it\widetilde M}$.
Starting from the same pure state $|\psi\rangle$, we compare the final states $U(t)|\psi\rangle$ and $\widetilde U(t)|\psi\rangle$.  
The loss of their overlap measures the operational visibility introduced above. 
It vanishes when the two orderings produce the same state up to a phase, and increases as their dynamical effects become distinguishable.

This construction deliberately isolates the ordering structure.  
At this stage, our goal is to establish whether the proposed mechanism can exist.  
For this purpose, we need two Hermitian generators that contain the same factors in opposite orders, together with input states on which their operations can be compared.  
We can therefore test the mechanism in a controlled finite system before introducing the theory-specific structure of a non-Abelian lattice gauge theory.  
A natural next step is to construct the corresponding operations from actual plaquette terms and test whether the physical states of a non-Abelian gauge theory suppress their ordering distinction.  
We leave this gauge-theory implementation to future work.

Comparing the overlap of states produced by two evolutions is familiar from quantum-state fidelity and Loschmidt echoes~\cite{Peres1984,Jalabert2001,Gorin2006,Quan2006,Wang2009,Jozsa1994,Zanardi2006}.
Out-of-time-order correlators also compare different operator orders, but are usually used to study how operators change with time~\cite{Maldacena2016,Swingle2018,Hashimoto2017}.
These comparisons provide tools for distinguishing operations through their output states.  
Our use of them is specific.  
We do not measure a commutator or a ``non-Abelian observable''.  
We send the same input state through two operations generated by opposite orderings of the same factors and ask whether their outputs can be distinguished.  
Measurements enter only as a way to distinguish those output states.  
We then change the input state while keeping the two operations fixed.  
The central object of this work is instead the operation-level question of whether non-Abelian ordering information can become operationally invisible.  
Abelian dominance motivates this question, but is not the conclusion established by the present test.

We first formulate this operation-discrimination test and show that the output overlap bounds the ability of any measurement to distinguish the two operations on the chosen input state.  
We then solve the test exactly for one qubit and demonstrate the same mechanism in a finite many-body system.  
In both cases, we keep the two operations fixed and change only the input state.  
These results establish that the proposed mechanism can occur. 
The present calculation does not implement the corresponding plaquette operations in a non-Abelian gauge theory. 
Testing whether its physical states realize this mechanism is a question for future work.

The paper follows this question in four steps.
Sec.~\ref{sec:definition} constructs the ordering test, gives its operational meaning, removes an arbitrary common time scale, and demonstrates the state dependence exactly for one qubit.  
Sec.~\ref{sec:short} then identifies the state-dependent quantities that control the short-time behavior.  
Sec.~\ref{sec:model} defines the many-body states and the fixed ordering test, and Sec.~\ref{sec:results} presents the endpoint contrast, the additional operator pairs, the intermediate states, and the coefficient-change tests.  
The final summary collects the conclusions.

\section{A state-dependent test of operator ordering}
\label{sec:definition}

How can the ordering information identified in the previous section be tested?  
The two ordered generators may be different as operators, but the mechanism concerns whether the operations they generate can be distinguished on a chosen input state.  
We must therefore compare their output states rather than measure a commutator.  
This section constructs that comparison, gives it a direct operational meaning, and specifies the required time window.  
It then shows the state dependence exactly in a one-qubit example and identifies its general short-time origin.

\subsection{\label{sec:2.1}The ordering test and its operational meaning}

The unitary evolutions generated by $A_1A_2$ and $A_2A_1$ are compared.  
The generators contain the same two factors in opposite orders.
Starting from the state that is actually prepared, we ask whether the two operations produce distinguishable final states.

Let $B=B^\dagger$ and $C=C^\dagger$ act on a finite-dimensional Hilbert space.
We define
\begin{equation}
A_1=B+iC,\qquad A_2=B-iC=A_1^\dagger,
\label{eq:Adefs}
\end{equation}
and the two ordered generators,
\begin{align}
M&=A_1A_2=B^2+C^2-i[B,C],\nonumber\\
\widetilde M&=A_2A_1=B^2+C^2+i[B,C].
\label{eq:orderedgenerators}
\end{align}
Both are Hermitian.  
Their expectation values are nonnegative because they are all in the form of $AA^\dagger$ or $A^\dagger A$.  
For a normalized pure state $|\psi\rangle$, define
\begin{equation}
\mathcal A_\psi(\tau;B,C)=
\langle\psi|e^{+i\widetilde M\tau}e^{-iM\tau}|\psi\rangle,
\label{eq:amplitude}
\end{equation}
where $\tau$ has units inverse to those of $M$.  
We use the magnitude of this overlap,
\begin{equation}
F_\psi(\tau;B,C)=|\mathcal A_\psi(\tau;B,C)|.
\label{eq:echo}
\end{equation}
It compares the two states $e^{-iM\tau}|\psi\rangle$ and $e^{-i\widetilde M\tau}|\psi\rangle$.  
Unitarity gives $0\leq F_\psi\leq1$.

The overlap has a direct measurement meaning.  
Using
\begin{equation}
\rho_M=|\psi_M\rangle\langle\psi_M|,\quad
\rho_{\widetilde M}=|\psi_{\widetilde M}\rangle
\langle\psi_{\widetilde M}|,
\end{equation}
where $|\psi_M\rangle=e^{-iM\tau}|\psi\rangle$ and $|\psi_{\widetilde M}\rangle=e^{-i\widetilde M\tau}|\psi\rangle$.  
Their trace distance is,
\begin{equation}
\frac12\|\rho_M-\rho_{\widetilde M}\|_1=\sqrt{1-F_\psi(\tau;B,C)^2}.
\label{eq:tracedistance}
\end{equation}
Therefore, for any observable $O$,
\begin{equation}
\left|\operatorname{Tr}\!\left[O(\rho_M-\rho_{\widetilde M})\right]\right| \leq 2\|O\|_{\rm op}\sqrt{1-F_\psi^2}.
\label{eq:observablebound}
\end{equation}
Here $\|O\|_{\rm op}$ is the largest absolute value of the eigenvalues of $O$.  
When $F_\psi$ is close to one, the two ordered operations give nearly the same expectation value for any observable $O$.
Suppose that the relevant non-Abelian distinction in a measurement is carried by the difference between these two ordered operations.
Eq.~\eqref{eq:observablebound} then guarantees that hiding this difference is sufficient for an Abelian-like measurement outcome.  
It is not necessary, because the same outcome may arise through a different mechanism.  
If $[B,C]=0$, then $M=\widetilde M$ and $F_\psi=1$ for every state and every time.
However, the converse is false, $F_\psi=1$ for a particular state or interval does not imply $[B,C]=0$.  
Low visibility means that this state does not reveal the ordering difference in this test, while the microscopic operators remain non-commuting.

The ordering test applies to both pure and mixed input states.  
A finite-temperature state is represented by a density matrix $\rho$, while a pure ground state is the special case $\rho=|\psi\rangle\langle\psi|$.  
The operations being compared and the meaning of their real-time visibility are unchanged; only the input state differs.  
We use pure ground states below to isolate how the visibility changes across a quantum phase transition, but the state dependence itself is not restricted to zero temperature.

The real-time construction suggests a possible extension to imaginary time.  
Formally, the product $R(\tau)=e^{-iM\tau}e^{+i\widetilde M\tau}$ continues under $i\tau\mapsto\beta$ to $R_{\rm E}(\beta)=e^{-\beta M}e^{+\beta\widetilde M}$, which involves the same two ordered generators and resembles the transfer-matrix structure used in quantum--classical descriptions of thermal transitions \cite{Wegner1971,Kogut1979,FradkinSusskind1978}.  
However, $R_{\rm E}(\beta)$ is non-unitary and is not itself a thermal density matrix.  
Defining a normalized imaginary-time visibility and determining its relation to thermal observables require additional choices and are beyond the scope of the present work.

\subsection{\label{sec2.2}Scale-independent visibility over a finite time window}

A common rescaling $B\mapsto sB$ and $C\mapsto sC$ changes both generators by $s^2$. 
It merely changes the clock used in Eq.~\eqref{eq:amplitude}.  
To remove this arbitrary common scale, define
\begin{equation}
\Omega(B,C)=\|M\|_{\rm op},\qquad u=\Omega\tau,
\label{eq:scale}
\end{equation}
where $\|\cdot\|_{\rm op}$ is the largest singular value.  
Since $M=A_1A_1^\dagger$ and $\widetilde M=A_1^\dagger A_1$ have the same nonzero spectrum, either ordering gives the same $\Omega$.  
For a nontrivial pair, $\Omega>0$, the identically zero pair has $F=1$ and is excluded from the rescaling.

The purpose of this rescaling is to remove the arbitrary time scale produced by a common change in the magnitude of $B$ and $C$.  
At fixed $u$, the comparison is unchanged under $B\mapsto sB$ and $C\mapsto sC$.  
It does not remove the dependence on the dimensionless window $U$, which specifies how long the two operations are compared.

We write $\widehat F_\psi(u)=F_\psi(u/\Omega)$ and choose a time window $0\leq u\leq U$.  
The visibility up to $U$ is
\begin{equation}
\mathcal V_\psi(U;B,C)=1-\min_{0\leq u\leq U}\widehat F_\psi(u;B,C).
\label{eq:visibility}
\end{equation}
It is zero when the ordering difference is invisible in the chosen window. 
It increases when the state permits a deeper loss of overlap.  
The window cannot be omitted. 
The two final states can be nearly identical at the beginning and different later.  
To check that a result is not caused by one narrow minimum, we also use the average visibility
\begin{equation}
\overline{\mathcal V}_\psi(U)=1-\frac{1}{U}\int_0^U\widehat F_\psi(u)\,du.
\label{eq:integrated}
\end{equation}
If $\mathcal V_\psi(U)\leq\epsilon$, then $F_\psi\geq1-\epsilon$ throughout the window.  
Eq.~\eqref{eq:observablebound} then bounds every measurement difference in that window by $2\|O\|_{\rm op}\sqrt{2\epsilon-\epsilon^2}$.  
The visibility thus measures the largest opportunity within the window to reveal the operator order.
The visibility in Eq.~\eqref{eq:visibility} cannot decrease when $U$ is increased. 
The average visibility may decrease.  Both lie between zero and one and are invariant under a common unitary change of basis applied to $B$, $C$, and the state.  
They are not invariant under independently changing the relative coefficients in $B$ and $C$. 
This change defines a different operator pair and a different test.

\subsection{A one-qubit example}
\label{sec:qubit}

\begin{figure}[t]
\centering
\includegraphics[width=0.98\hsize]{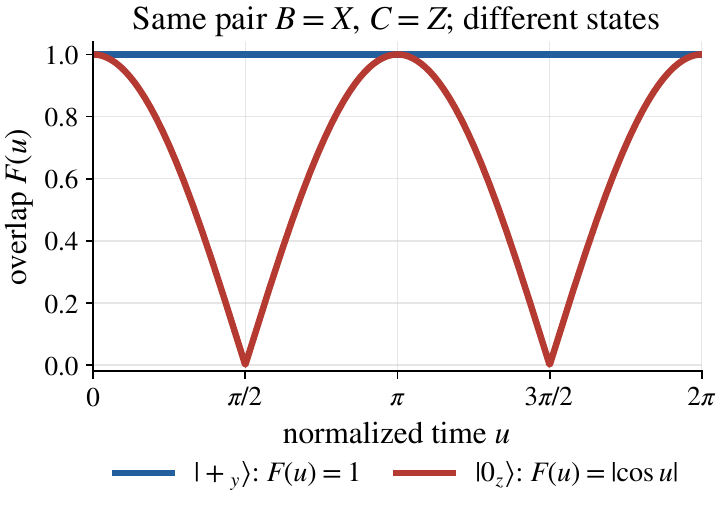}
\caption{\label{fig:qubit}
State dependence for a fixed one-qubit ordering test.  
The pair $B=X$, $C=Z$ and hence the two ordered generators are identical for both curves.  
In the $Y$ eigenstate $|+_y\rangle$, reversing the order changes only the phase and $F(u)=1$ for all $u$.  
In the $Z$ eigenstate $|0_z\rangle$, the same ordering difference is operationally visible, with $F(u)=|\cos u|$.}
\end{figure}

The state dependence is already visible with one qubit.  
Set $B=bX$ and $C=cZ$, where $X$ and $Z$ are Pauli matrices.  
Since $[X,Z]=-2iY$~($Y$ is also the Pauli matrix), Eq.~\eqref{eq:orderedgenerators} gives
\begin{equation}
M=(b^2+c^2)\mathbb I-2bcY,\qquad
\widetilde M=(b^2+c^2)\mathbb I+2bcY.
\end{equation}
The two generators commute with each other, so the common scalar part cancels from the overlap and
\begin{equation}
F_\psi(\tau)=\left|\langle\psi|e^{i4bcY\tau}|\psi\rangle\right|.
\label{eq:qubitecho}
\end{equation}
For $b=c\neq0$, $\Omega=4b^2$ and $u=4b^2\tau$.  
A $Y$ eigenstate $|+_y\rangle$ has
\begin{equation}
F_{|+_y\rangle}(u)=1,
\end{equation}
whereas the $Z$ eigenstate $|0_z\rangle$ has
\begin{equation}
F_{|0_z\rangle}(u)=|\cos u|.
\end{equation}
The same pair has $[B,C]\neq0$ in both cases.  
The first state is an exact example of a state that hides the ordering difference from this test. 
The second makes it visible.  
Fig.~\ref{fig:qubit} displays the two curves.

\subsection{More general cases in the short time limit}
\label{sec:short}

Why can the same two ordered operations be hidden by one state and revealed by another?  
For general many-body operators $B$ and $C$, the exact finite-time overlap contains two different operator exponentials and is difficult to interpret directly.  
A short-time expansion provides a controlled analytic answer. 
It shows, order by order, which properties of the input state determine whether the two operations can be distinguished. 
Define,
\begin{equation}
D=B^2+C^2,\qquad K=\widetilde M-M=2i[B,C],
\label{eq:DK}
\end{equation}
so that $M=D-K/2$ and $\widetilde M=D+K/2$.  Expanding the two exponentials in Eq.~\eqref{eq:amplitude} gives,
\begin{equation}
F_\psi(\tau)=1-\frac{\tau^2}{2}
\left(\langle K^2\rangle_\psi-\langle K\rangle_\psi^2\right)
 +O(\tau^3).
\label{eq:shortleading}
\end{equation}
Here $\langle O\rangle_\psi=\langle\psi|O|\psi\rangle$.  
The first loss of overlap is therefore controlled by the state through $\operatorname{Var}_\psi(K)$.
For the same fixed $B$ and $C$, different fluctuations of $K$ give different initial rates at which the two operations become distinguishable.  
This is the leading analytic origin of state-dependent visibility.  
It is not the complete finite-time answer, because higher orders contain expectation values of longer ordered products of $D$ and $K$.

To identify the state dependence beyond the leading term, expand the amplitude as,
\begin{equation}
\mathcal A_\psi(\tau)=\sum_{n\geq0}a_n\tau^n.
\end{equation}
Direct multiplication of the two exponential series gives the coefficient recurrence,
\begin{equation}
a_n=\sum_{r=0}^{n}\frac{i^r(-i)^{n-r}}{r!(n-r)!}
\left\langle\widetilde M^rM^{n-r}\right\rangle_\psi.
\label{eq:an}
\end{equation}
If $R=|\mathcal A|^2=\sum_n r_n\tau^n$, then $r_n=\sum_{m=0}^na_ma_{n-m}^*$.  
Writing $F=\sum_nf_n\tau^n$ with $f_0=1$, the coefficients follow from,
\begin{equation}
f_n=\frac12\left(r_n-\sum_{m=1}^{n-1}f_mf_{n-m}\right),\qquad n\geq1.
\label{eq:fn}
\end{equation}
Eqs.~\eqref{eq:an} and \eqref{eq:fn} give every coefficient directly.

The expansion separates three different statements.
First, $[B,C]\neq0$ is an operator statement, the two orderings define different generators, independently of the state.  
Second, $\operatorname{Var}_\psi(K)$ determines the leading quadratic loss of overlap and therefore how rapidly the ordering difference first becomes visible in the state $|\psi\rangle$.  
Third, the finite-time visibility also depends on higher ordered moments such as $\langle\widetilde M^rM^{n-r}\rangle_\psi$.  
Two states can consequently have the same initial quadratic loss and still distinguish the two operations very differently at later times.  
The commutator identifies the ordering difference. 
The state determines whether and when that difference becomes visible.

\section{A finite many-body test}
\label{sec:model}

\subsection{States and lattice}

Can the same mechanism occur beyond one qubit?  
To answer this question, we need a controlled family of many-body states while keeping the ordering test fixed.  
We obtain such states from a finite $\mathbb Z_2$ lattice Hamiltonian, following the Hamiltonian formulation of lattice gauge theory~\cite{Wilson1974,Kogut1979,Wegner1971,KogutSusskind1975,FradkinSusskind1978}.

\begin{figure*}[t]
\centering
\includegraphics[width=0.96\hsize]{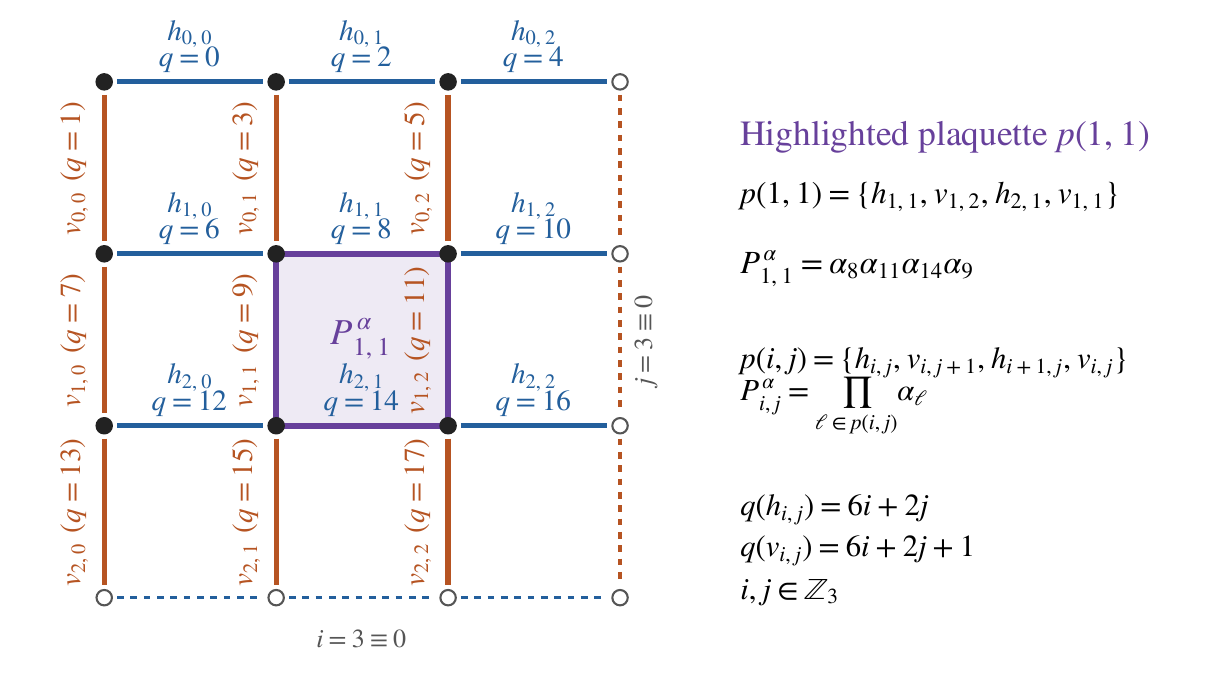}
\caption{\label{fig:lattice}
Lattice geometry, plaquette orientation, and qubit-index conventions.  
Horizontal links $h_{i,j}$ are blue and carry $q(h_{i,j})=6i+2j$; vertical links $v_{i,j}$ are orange and carry $q(v_{i,j})=6i+2j+1$, with $i,j\in\mathbb Z_3$.  
The highlighted central cell gives the explicit example $p(1,1)=\{h_{1,1},v_{1,2},h_{2,1},v_{1,1}\}$, corresponding to qubits $\{8,11,14,9\}$, so that $P_{1,1}^{\alpha}=\alpha_8\alpha_{11}\alpha_{14}\alpha_9$.  
The formulas at right give the corresponding definitions.  
Open sites and dashed links on the right and lower boundaries are periodic copies and introduce no additional qubits.}
\end{figure*}

The square lattice has $3\times3$ sites with periodic boundary conditions and eighteen link qubits.  
Figure~\ref{fig:lattice} displays the lattice geometry, the link labels, the qubit-index convention, and the four links entering one plaquette operator.

For each plaquette $p$, let $P_p^Z=\prod_{\ell\in p}Z_\ell$.  
We diagonalize
\begin{equation}
H(\lambda)=-\sum_pP_p^Z-\lambda\sum_\ell X_\ell
\label{eq:H}
\end{equation}
at
\begin{equation}
\lambda\in\{0.05,0.10,0.20,0.40,0.80,1.60,3.20\}.
\label{eq:lambdas}
\end{equation}
For the same $3\times3$ periodic $\mathbb Z_2$ lattice with eighteen link qubits, it has been found that the second derivative of the ground-state energy reaches its minimum at $g\simeq0.380$~\cite{Cui:2019sfz}.
This is why our sampled values include $\lambda=0.40$ at the middle.
The normalized lowest-energy eigenvector is denoted by $|\psi(\lambda)\rangle$.  
The index of qubits corresponding to the links are shown in Fig.~\ref{fig:lattice}.

Changing $\lambda$ gives a reproducible family of many-body states.  
We use the same $B$ and $C$ in every state, so any change in the ordering visibility comes from the state.  
This lets the finite model demonstrate the mechanism beyond the exactly solvable one-qubit example.

\subsection{The fixed operator pair}
\label{sec:pairsearch}

\begin{figure*}[t]
\centering
\includegraphics[width=0.92\hsize]
{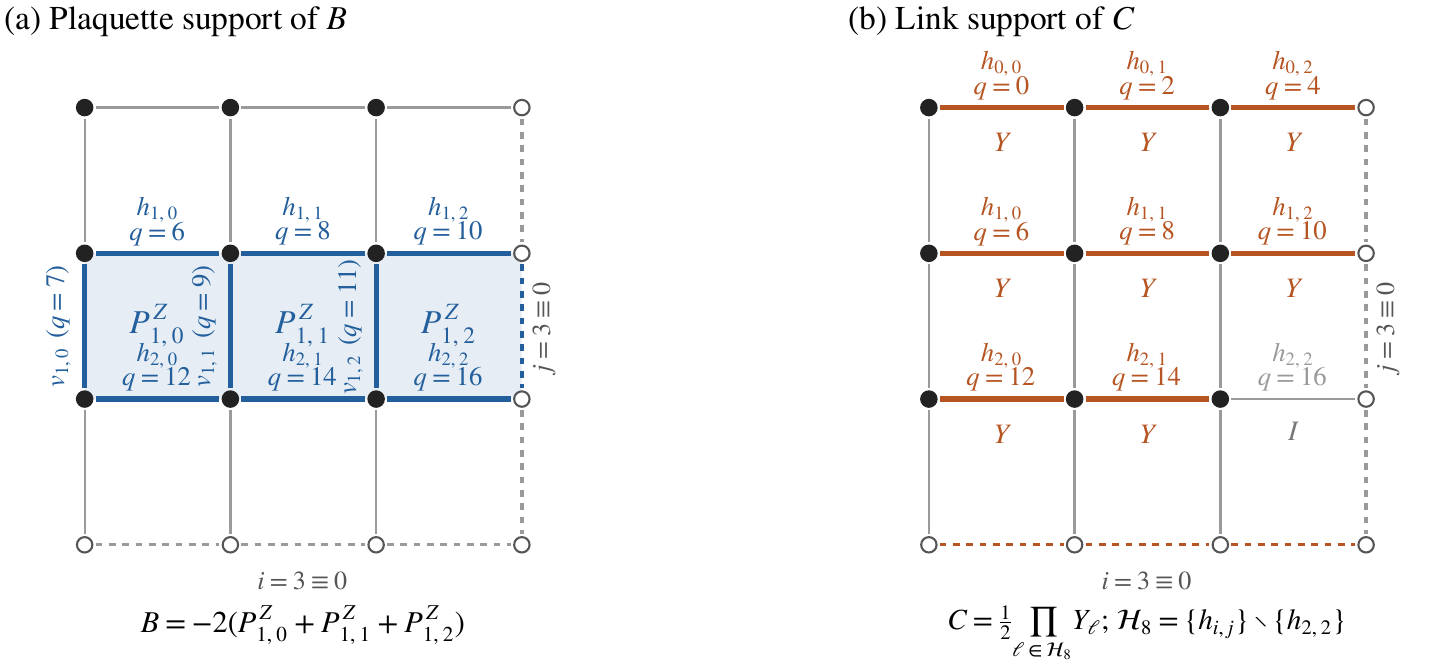}
\caption{\label{fig:geometry}
Spatial support of the fixed operator pair in Eq.~\eqref{eq:BC}.  
(a) The three shaded plaquettes are $P^Z_{1,0}$, $P^Z_{1,1}$, and $P^Z_{1,2}$ in $B=-2\sum_jP^Z_{1,j}$. 
Each plaquette term applies $Z$ to its four boundary links.  
Link labels include the corresponding qubit index $q$.  
(b) The product $C$ applies $Y$ to the eight orange horizontal links in $\mathcal H_8$.  
The only horizontal link excluded from that product is $h_{2,2}$, or $q=16$, on which $C$ acts as the identity.}
\end{figure*}

For a plaquette $p(i,j)$ define
\begin{equation}
P_{i,j}^{\alpha}=\prod_{\ell\in p(i,j)}\alpha_\ell,
\qquad \alpha\in\{X,Y,Z\}.
\end{equation}
Let
\begin{equation}
\mathcal H_8=\{h_{0,0},h_{0,1},h_{0,2},h_{1,0},h_{1,1},h_{1,2},h_{2,0},h_{2,1}\}.
\end{equation}
The main pair used in every state comparison is
\begin{align}
B&=-2\left(P^Z_{1,0}+P^Z_{1,1}+P^Z_{1,2}\right),\nonumber\\
C&=\frac12\prod_{\ell\in\mathcal H_8}Y_\ell.
\label{eq:BC}
\end{align}
The pair acts nontrivially on twelve of the eighteen links and has $[B,C]\neq0$.  
Its common normalization is
\begin{equation}
\Omega=\|M\|_{\rm op}=36.3745155.
\label{eq:omega}
\end{equation}
where $\Omega$
is defined in Eq.~\eqref{eq:scale}.
Fig.~\ref{fig:geometry} unpacks Eq.~\eqref{eq:BC}.  
The left panel shows the three plaquette products added in $B$.  
The right panel lists the nine horizontal links and marks the eight on which the product $C$ contains $Y$. 
The remaining link contributes the identity.

We found this pair through an LLM-agent-guided adaptive search inspired by FunSearch~\cite{RomeraParedes2024,Song:2025pwy,Cao:2025shc,Cao:2025fla}.  
The target was not merely to maximize the difference between the two states.  
We sought a pair whose ordering difference is nearly invisible in $|\psi(0.05)\rangle$ and clearly visible in $|\psi(3.20)\rangle$.  
For each candidate pair, define
\begin{equation}
m_\lambda=\min_{\tau\in\mathcal T}F_{\psi(\lambda)}(\tau),
\qquad \mathcal T=\{0.1,0.2,\ldots,1.0\},\
\label{eq:searchminimum}
\end{equation}
where $F_{\psi(\lambda)}(\tau)$ denotes
$F_{\psi(\lambda)}(\tau;B,C)$ from Eq.~\eqref{eq:echo}, evaluated for the
candidate pair $B,C$.  The search maximized
\begin{equation}
S=m_{0.05}\bigl(m_{0.05}-m_{3.20}\bigr).
\label{eq:searchscore}
\end{equation}
Since $1-m_\lambda$ is the largest loss of overlap on the sampled grid, the second factor rewards a large visibility contrast.  
The prefactor rewards $m_{0.05}\simeq1$ and therefore penalizes candidates whose ordering difference is already visible in $|\psi(0.05)\rangle$.  
A high score thus requires both properties.  
We separately asked the agent to construct operators with clear geometric interpretations.
It wrote and iteratively revised the search programs.  
After selection, we reevaluated the pair using the scale-independent time $u$ and the visibility in Eq.~\eqref{eq:visibility}.
The results first compare the two endpoint states.  
We then ask whether the same contrast persists for other operator geometries, how it develops through the intermediate states, and whether it survives small coefficient changes.

\section{Results: the state changes what the same ordering test can see}
\label{sec:results}

\subsection{A fixed pair gives two different answers}

\begin{figure*}[t]
\centering
\includegraphics[width=0.95\hsize]{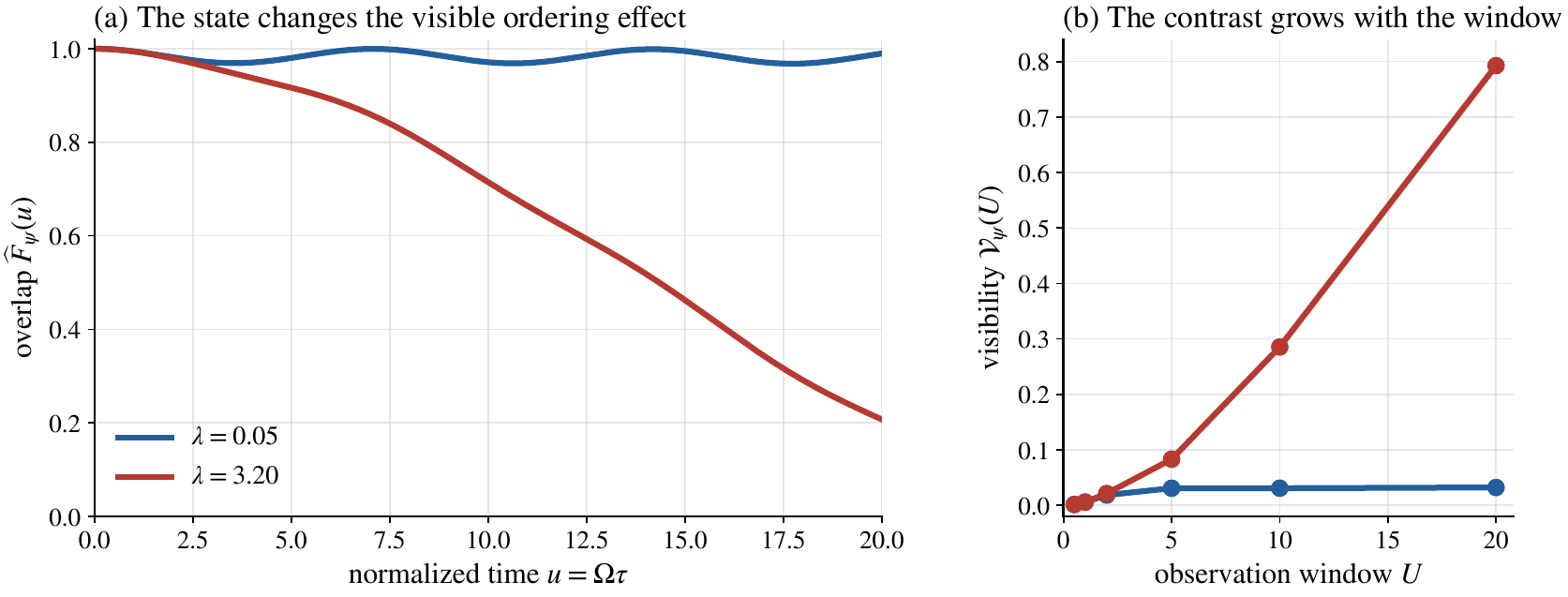}
\caption{\label{fig:mainresult}
State-dependent visibility for the fixed pair in Eq.~\eqref{eq:BC}.  
The operators $B,C$ and their normalization $\Omega$ are held fixed; at each $U$, the same observation window is used for both states, so only the input ground state $|\psi(\lambda)\rangle$ changes.  
(a) The normalized overlap $\widehat F_{\psi(\lambda)}(u)$ stays close to one for $\lambda=0.05$ but develops deep minima for $\lambda=3.20$.  
(b) The corresponding window visibility $\mathcal V_{\psi(\lambda)}(U)=1-\min_{0\leq u\leq U}\widehat F_{\psi(\lambda)}(u)$ separates as $U$ grows, reaching $0.032098$ and $0.792679$, respectively, at $U=20$.}
\end{figure*}

We begin with the central question. Does changing only the state affect the visibility of a fixed non-commuting pair?  
Fig.~\ref{fig:mainresult}(a) shows the numerically evaluated normalized overlaps for $\lambda=0.05$ and $\lambda=3.20$.
They start together, but separate strongly at later times.  
At $U=20$,
\begin{equation}
\mathcal V_{0.05}=0.032098,\qquad
\mathcal V_{3.20}=0.792679,
\label{eq:mainnumbers}
\end{equation}
for the same $B$, $C$, normalization, and window. 
Therefore, the difference does not arise from changing the operators. 
It is also not caused by changing their overall scale. 

Could this contrast be caused by choosing one favorable time?  
The answer is no.  
Fig.~\ref{fig:mainresult}(b) reports the minimum over each complete window.  
The visibility contrast $\Delta\mathcal V=\mathcal V_{3.20}-\mathcal V_{0.05}$ is only $0.000012$ at $U=0.5$, grows to $0.052240$ at $U=5$, and reaches $0.760581$ at $U=20$.
The average visibilities at the final window are
\begin{equation}
\overline{\mathcal V}_{0.05}=0.016776,\qquad
\overline{\mathcal V}_{3.20}=0.321993.
\end{equation}
Thus the result is not produced by a single isolated dip in one curve.

\begin{table}[t]
\caption{\label{tab:windows}
Visibility up to each value of $U$ for the two endpoint states.  
The operators and normalization are identical in every row.}
\begin{ruledtabular}
\begin{tabular}{cccc}
$U$ & $\mathcal V_{0.05}$ & $\mathcal V_{3.20}$ & $\Delta\mathcal V$\\
\hline
0.5 & 0.001487 & 0.001499 & 0.000012\\
1 & 0.005661 & 0.005849 & 0.000188\\
2 & 0.018482 & 0.021262 & 0.002780\\
5 & 0.030813 & 0.083053 & 0.052240\\
10 & 0.030813 & 0.285423 & 0.254610\\
20 & 0.032098 & 0.792679 & 0.760581\\
\end{tabular}
\end{ruledtabular}
\end{table}

It can be seen from Fig.~\ref{fig:mainresult} that the curves first agree and then separate, which can be explained by the short-time expansion.
For both endpoint states,
\begin{equation}
\operatorname{Var}_{\psi(0.05)}(K)=\operatorname{Var}_{\psi(3.20)}(K)=16,
\end{equation}
where $\operatorname{Var}(K)$ is the leading-order coefficient in Eq.~\eqref{eq:shortleading}.
The odd coefficients vanish within numerical precision through the computed order.  
Keeping the first two resolved nonzero terms gives
\begin{align}
\widehat F_{0.05}(u)&\simeq1-0.0060463816u^2+0.0003956658u^4,\nonumber\\
\widehat F_{3.20}(u)&\simeq1-0.0060463816u^2+0.0002024636u^4.
\end{align}

\begin{figure*}[t]
\centering
\includegraphics[width=0.92\hsize]{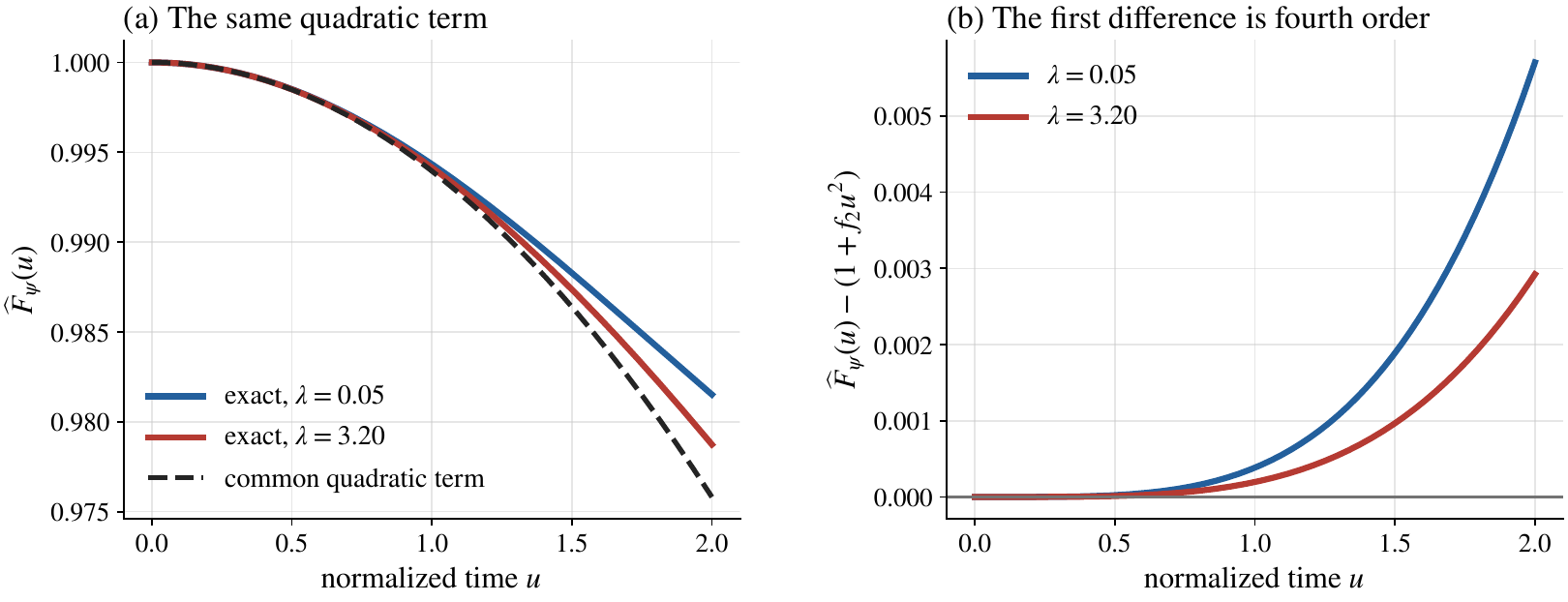}
\caption{\label{fig:short}
Short-time origin of the endpoint contrast, using $u=\Omega\tau$.  
(a) The exact overlaps for $\lambda=0.05$ and $3.20$ share the dashed quadratic approximation because both states have $\operatorname{Var}_{\psi}(K)=16$.  
(b) Subtracting this common quadratic term exposes different residuals.  Their first numerically resolved difference is the fourth-order contribution, so equal commutator variance does not determine the finite-time visibility.}
\end{figure*}
The fourth-order term is therefore the first state-dependent coefficient resolved by the calculation.  
Their common quadratic decrease is fixed by the commutator variance.  
Their later separation depends on longer ordered products of $D$ and $K$, as predicted in Sec.~\ref{sec:short}.

\subsection{The result is not confined to the main pair}

\begin{table*}[t]
\centering
\scriptsize
\setlength{\tabcolsep}{4pt}
\caption{\label{tab:multiplepairs}Definitions and $U=20$ visibilities of the
nine pairs in Fig.~\ref{fig:multiplepairs}.  The operators $P^\alpha_{i,j}$,
$\Pi_\alpha$, $\Sigma_\alpha$, $R_Z$, $L_Z$, and $Q_\alpha^\pm$ are defined in
Eqs.~\eqref{eq:BC}, \eqref{eq:pairnotation}, and \eqref{eq:pairshorthand}.
Here $k$ is the number of links on which $B$ or $C$ is not the identity.}
\begin{ruledtabular}
\begin{tabular}{cllccc}
pair & $B$ & $C$ & $k$ & $\mathcal V_{0.05}$ & $\mathcal V_{3.20}$\\
\hline
A & $R_Z$ & $\tfrac12\Pi_Y(\{0,2,4,6,8,10,12,14\})$ & 12 & 0.032098 & 0.792679\\
B & $L_Z$ & $-\tfrac12P^Y_{0,0}$ & 12 & 0.032098 & 0.792679\\
C & $L_Z$ & $\tfrac12\Pi_X(\{1,7,13,17\})$ & 12 & 0.032098 & 0.792611\\
D & $L_Z$ & $-\tfrac12\Pi_Y(\{1,3,5,7,13,15\})$ & 12 & 0.032098 & 0.792679\\
E & $Q_Y^+$ & $\tfrac12\Sigma_Z(\{0,2,4,8,14\})$ & 15 & 0.033224 & 0.830820\\
F & $Q_Z^-$ & $\tfrac12\Sigma_X(\{0,6,8,12,16\})$ & 13 & 0.067543 & 0.478966\\
G & $Q_Z^+$ & $-\tfrac12\Sigma_X(\{1,3,13\})$ & 12 & 0.072082 & 0.721102\\
H & $Q_Z^+$ & $\tfrac12\Sigma_Y(\{1,13,15,17\})$ & 12 & 0.083469 & 0.883209\\
I & $R_Z$ & $-\tfrac12\Pi_X(\{0,2,4,6,10,12,14,16\})$ & 12 & 0.032077 & 0.792609\\
\end{tabular}
\end{ruledtabular}
\end{table*}
The first comparison shows that the effect can occur. 
We then ask whether it depends on an unusually chosen geometry. 
The FunSearch-inspired program produced many candidate pairs.  
From them, we selected the ten distinct pairs with the highest values of the search score in Eq.~\eqref{eq:searchscore}.  
Two of the ten are related by a translation, rotation, or reflection of the lattice.  
Keeping one representative of this symmetry-related pair leaves nine inequivalent pairs, which we label A--I.  
For an explicit definition of every pair, let
\begin{equation}
\Pi_\alpha(S)=\prod_{q\in S}\alpha_q,
\qquad
\Sigma_\alpha(S)=\sum_{q\in S}\alpha_q,
\label{eq:pairnotation}
\end{equation}
where $S$ is a set of link indices and $\alpha\in\{X,Y,Z\}$.  
We also define
\begin{align}
R_Z&=-2\sum_{j=0}^{2}P^Z_{1,j},\nonumber\\
L_Z&=-2\sum_{i=0}^{2}P^Z_{i,1},\nonumber\\
Q_\alpha^{\pm}&=\pm2\left(P^\alpha_{0,0}+P^\alpha_{2,0}
+P^\alpha_{0,2}+P^\alpha_{2,2}\right).
\label{eq:pairshorthand}
\end{align}
Table~\ref{tab:multiplepairs} then gives the complete definitions and the displayed values.  
Pair A is the main pair in Eq.~\eqref{eq:BC}.
Pairs A and I share $R_Z$ but use different link products.
Pairs E--H use corner plaquettes and single-link sums.
Every pair is evaluated in the same two endpoint states, with its own common scale $\Omega=\|M\|_{\rm op}$ and the same window $U=20$.

\begin{figure*}[t]
\centering
\includegraphics[width=0.95\hsize]{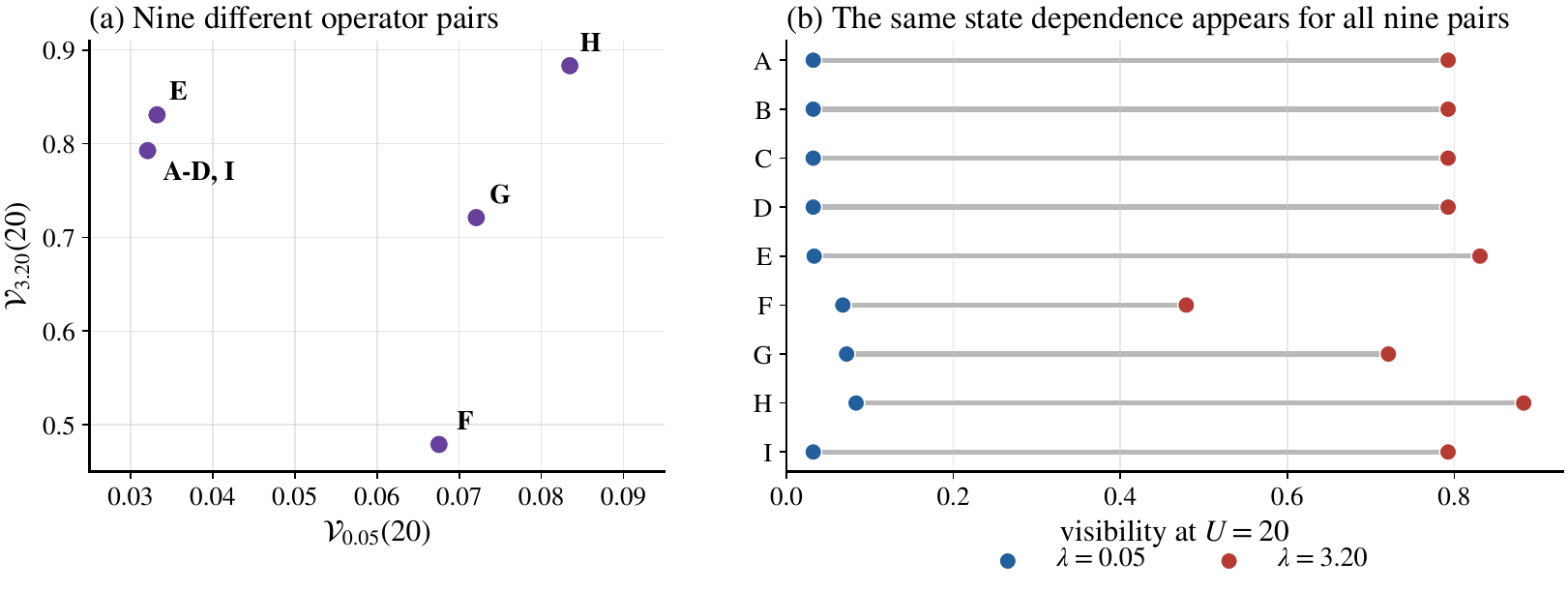}
\caption{\label{fig:multiplepairs}
A comparison for the nine search-selected, symmetry-inequivalent pairs A--I defined in Table~\ref{tab:multiplepairs}.  
These pairs are selected outputs of the search in Sec.~\ref{sec:pairsearch}, not a random sample.  
(a) Each point has coordinates $[\mathcal V_{0.05}(20),\mathcal V_{3.20}(20)]$. 
Coincident labels denote values equal to the displayed precision.  
(b) Paired markers show the same two visibilities for each operator pair.  
Every pair gives a substantially larger visibility in $|\psi(3.20)\rangle$ than in $|\psi(0.05)\rangle$.}
\end{figure*}
Fig.~\ref{fig:multiplepairs} shows all nine results.  
In the $\lambda=0.05$ state their visibilities lie between $0.0321$ and $0.0835$, whereas in the $\lambda=3.20$ state they lie between $0.4790$ and $0.8832$.
The visibility contrast therefore ranges from $0.4114$ to $0.7997$.  
The set includes two paired plaquettes. 
It also includes products along several links and sums of single-link Pauli operators. 
Therefore, the effect is not limited to the geometry of pair A.

\subsection{The intermediate states connect the two endpoints}

\begin{figure*}[t]
\centering
\includegraphics[width=0.95\hsize]{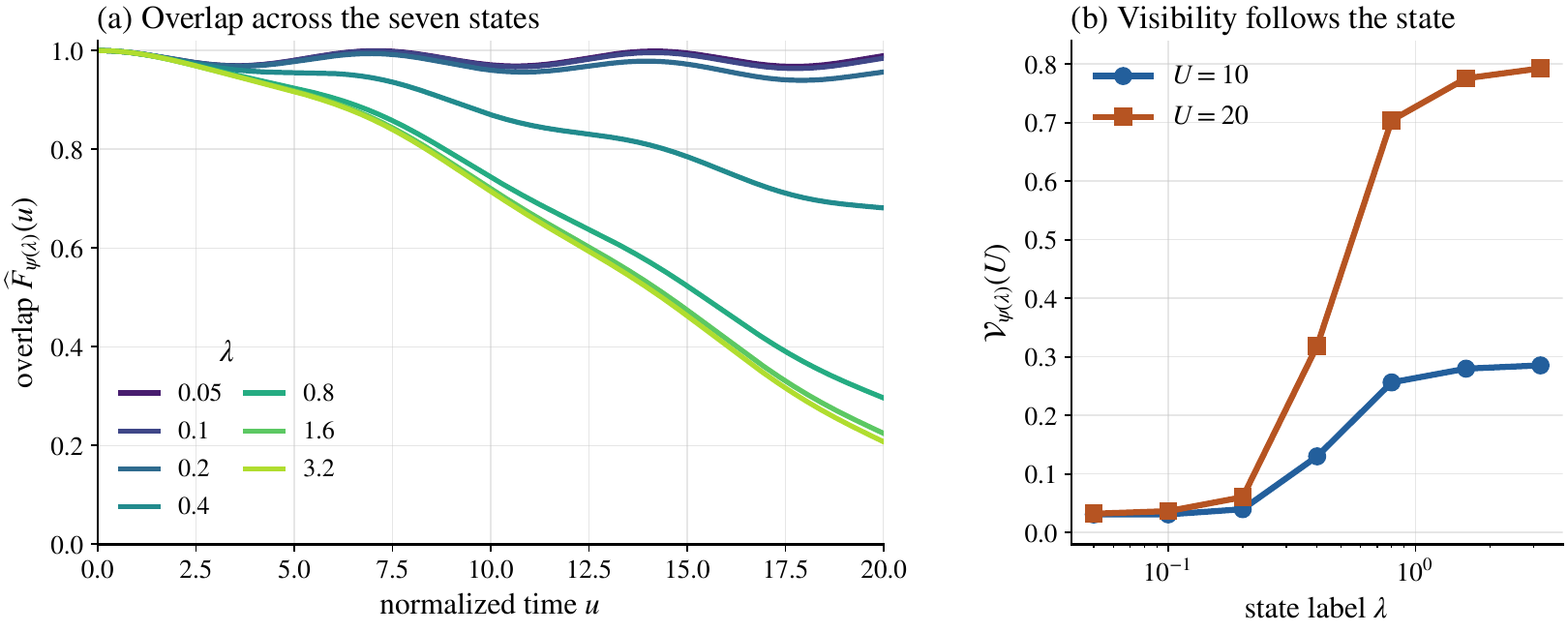}
\caption{\label{fig:family}
Evolution through the seven-state family~($0.05\leq \lambda\leq 3.20$) for the fixed pair in Eq.~\eqref{eq:BC}. 
(a) The curves are the normalized overlaps $\widehat F_{\psi(\lambda)}(u)$. 
Neither the operators nor their normalization changes with $\lambda$. 
(b) The visibility at $U=10$ and $20$ remains small for $\lambda=0.05$--$0.20$, increases through the finite-size transition region near $\lambda=0.40$, and is large on the high-$\lambda$ side.}
\end{figure*}
The two endpoint states show that the effect can occur.  
To check what happens between them, we evaluate the same pair in all seven states $|\psi(\lambda)\rangle$ where the values $\lambda$ are taken from Eq.~\eqref{eq:lambdas}. 
Fig.~\ref{fig:family}(a) displays the overlap for all seven states, and Fig.~\ref{fig:family}(b) displays visibility at two fixed windows.  
The visibility changes across the sampled values of $\lambda$. 
At $U=20$ it rises from $0.0321$ at $0.05$ to $0.7927$ at $3.20$, with the intermediate states filling the intervening behavior.  
The operator pair has not been changed.  
This supports state dependence within the seven-state family and connects the two endpoint examples.

Fig.~\ref{fig:family} shows more than an interpolation between the endpoints.  
At $U=20$, the visibility remains small from $\lambda=0.05$ to $0.20$, changing only from $0.0321$ to $0.0604$.  
It rises to $0.3189$ at $\lambda=0.40$ and reaches $0.7039$ at $\lambda=0.80$.  
Thus, for this fixed pair in the finite system, the two sides of the transition region have qualitatively different operational behavior.  
On the small-$\lambda$ side, changing the state leaves the ordering difference almost invisible.
Across the transition region near $\lambda=0.40$, the same ordering difference becomes clearly visible.  
The main change in visibility is therefore tied to the change of phase represented by this family of states.

\subsection{Small changes to the coefficients}

\begin{figure*}[t]
\centering
\includegraphics[width=0.88\hsize]{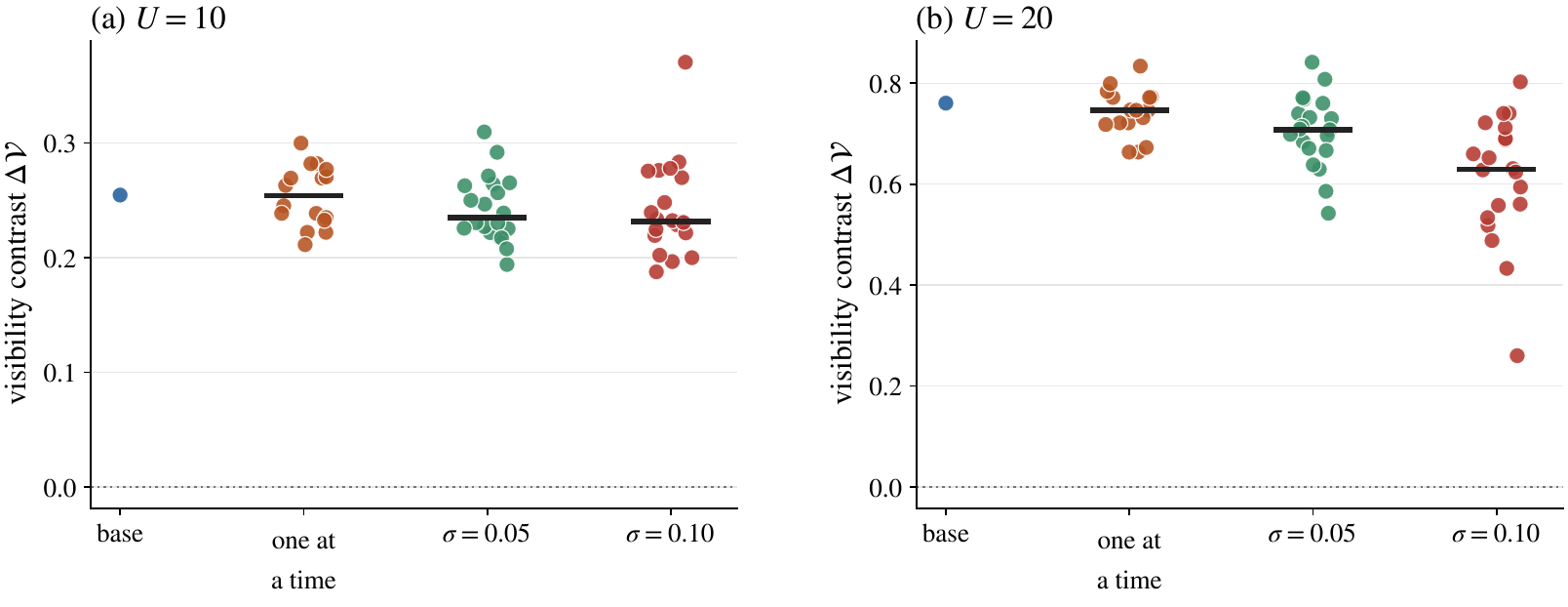}
\caption{\label{fig:robust}
Local coefficient robustness of the endpoint contrast $\Delta\mathcal V=\mathcal V_{3.20}-\mathcal V_{0.05}$ at $U=10$ (left) and $U=20$ (right).  
The 57 displayed cases comprise the unchanged pair, 16 one-at-a-time coefficient changes, and 20 simultaneous random changes at each of $\sigma=0.05$ and $0.10$; every case is retained.  
Each marker is one case, black horizontal segments are group medians, and the dotted line marks zero contrast.  
All tested values remain positive.}
\end{figure*}

\begin{table*}[t]
\caption{\label{tab:robustapp}Summary of all coefficient-change cases.
The entries are minimum, median, maximum, and mean of
$\Delta\mathcal V=\mathcal V_{3.20}-\mathcal V_{0.05}$.}
\begin{ruledtabular}
\begin{tabular}{lrrrrrrrrr}
group & $n$ & \multicolumn{4}{c}{$U=10$} & \multicolumn{4}{c}{$U=20$}\\
 & & min & median & max & mean & min & median & max & mean\\
\hline
one at a time & 16 & 0.211339 & 0.254140 & 0.299884 & 0.253679 & 0.663539 & 0.746374 & 0.833916 & 0.741618\\
$\sigma=0.05$ & 20 & 0.194094 & 0.234682 & 0.309654 & 0.242732 & 0.542432 & 0.708462 & 0.841593 & 0.704742\\
$\sigma=0.10$ & 20 & 0.187649 & 0.231623 & 0.370425 & 0.242452 & 0.260255 & 0.629577 & 0.802652 & 0.611835\\
\end{tabular}
\end{ruledtabular}
\end{table*}

Finally, we ask whether the contrast disappears under small changes of the four nonzero coefficients in Eq.~\eqref{eq:BC}.  
First, we multiply one coefficient at a time by $0.90$, $0.95$, $1.05$, or $1.10$, giving sixteen cases.  
We then change all four coefficients at once. 
Each is multiplied by $1+\epsilon$, where each $\epsilon$ is drawn independently from a normal distribution with mean zero and standard deviation $\sigma$.  
We use twenty draws at $\sigma=0.05$ and twenty at $\sigma=0.10$.  
Together with the unchanged pair, this gives $57$ cases.  
Every case was retained.
None was selected based on its visibility. 
The base contrast is $\Delta\mathcal V(10)=0.254610$ and $\Delta\mathcal V(20)=0.760581$.  
For the one-at-a-time changes the median contrasts are $0.254140$ and $0.746374$ at $U=10$ and $20$. 
For the $\sigma=0.05$ group they are $0.234682$ and $0.708462$.  
Even the broader $\sigma=0.10$ group has medians $0.231623$ and $0.629577$.  
Fig.~\ref{fig:robust} shows every case.  
The contrast is therefore locally stable under the tested coefficient changes.

\subsection{Summary of the numerical results}

The results establish the mechanism in two steps.  
The one-qubit example proves exactly that a state can hide the order of a fixed non-commuting pair.  
The finite lattice calculation then realizes the same effect in a family of many-body states.  
For the main pair, the visibility changes across the seven sampled states~($0.05\leq \lambda\leq 3.20$) and remains separated under every tested coefficient change.  
Eight additional inequivalent pairs show that the effect is not confined to the geometry of the main pair.

The exact limit of this mechanism reveals the special role of the initial state.  
And the operational visibility of non-commutativity provides a quantitative deviation from the exact limit.
Consider a normalized pure state for which
\begin{equation}
F_\psi(\tau)=1\qquad\text{for every }\tau.
\label{eq:exactinvisibility}
\end{equation}
With $W(\tau)=e^{+i\widetilde M\tau}e^{-iM\tau}$, Eq.~\eqref{eq:exactinvisibility} means that the same initial state is an eigenstate of every $W(\tau)$.  
Hence there is a real phase $\theta(\tau)$ such that
\begin{equation}
e^{-iM\tau}|\psi\rangle
=e^{i\theta(\tau)}e^{-i\widetilde M\tau}|\psi\rangle.
\label{eq:exactray}
\end{equation}
The phase disappears from the density matrix.  
Therefore the two orderings give exactly the same physical state and the same result for every observable at every time,
\begin{equation}
\rho_M(\tau)=\rho_{\widetilde M}(\tau),
\qquad
\operatorname{Tr}[O\rho_M(\tau)]=\operatorname{Tr}[O\rho_{\widetilde M}(\tau)].
\label{eq:exactoutputs}
\end{equation}

Eq.~\eqref{eq:exactinvisibility} also imposes a strict condition on the initial state itself.  
Let $K=\widetilde M-M$.  
Differentiating Eq.~\eqref{eq:exactray} shows that there is a fixed real number $\kappa$ for which
\begin{equation}
K e^{-i\widetilde M\tau}|\psi\rangle
=\kappa e^{-i\widetilde M\tau}|\psi\rangle
\qquad\text{for every }\tau.
\label{eq:trajectoryeigenstate}
\end{equation}
In particular, the initial state satisfies
\begin{equation}
K|\psi\rangle=\kappa|\psi\rangle,
\qquad \operatorname{Var}_\psi(K)=0,
\label{eq:sharpK}
\end{equation}
together with the full hierarchy
\begin{equation}
(K-\kappa)\widetilde M^n|\psi\rangle=0,
\qquad n=0,1,2,\ldots.
\label{eq:invisibilityhierarchy}
\end{equation}
Equivalently, define the centered ordering difference and its successive changes under the dynamics by
\begin{equation}
\begin{aligned}
\Delta K&=K-\langle K\rangle_\psi I,\\
C_1&=[\widetilde M,K],\qquad
C_{n+1}=[\widetilde M,C_n].
\end{aligned}
\label{eq:orderingdescendants}
\end{equation}
The exact invisible state then obeys the initial-state conditions
\begin{equation}
\Delta K|\psi\rangle=0,
\qquad
C_n|\psi\rangle=0,\qquad n\geq1.
\label{eq:initialstateconditions}
\end{equation}
Thus the initial state is an eigenstate of the ordering-difference operator $K=\widetilde M-M$ with eigenvalue $\kappa$, and every state $e^{-i\widetilde M\tau}|\psi\rangle$ along the trajectory remains in the same eigenspace of $K$ with the same eigenvalue.  Although $M$ and $\widetilde M$ remain different operators on the full Hilbert space, their difference acts as the constant $\kappa$ on this trajectory and therefore changes only the overall phase in Eq.~\eqref{eq:exactray}. 
Beyond gauge theory, the same visibility can diagnose whether reordering non-commuting gates in a product-formula quantum simulation creates a detectable error on the state being simulated, or whether alternative pulse orderings in coherent quantum control are distinguishable on a target state~\cite{Childs2021Trotter,Blanes2009Magnus}.

\section{Conclusion}
\label{sec:conclusion}

Where does the distinction between Abelian and non-Abelian multiplication enter the dynamics?  
In a lattice gauge theory, it enters through the order of the link variables in the Hamiltonian.  
We asked whether a state can hide the dynamical effect of this order even when the operators do not commute.  
To answer this question, we introduced the operational visibility of non-commutativity, i.e., the loss of overlap between two evolutions generated by opposite orderings of the same factors.  
We derived its measurement bound and short-time expansion, solved the test exactly for one qubit, and evaluated it in seven ground states of an eighteen-qubit $\mathbb Z_2$ lattice Hamiltonian.

Our main conclusion is that the state can determine whether a fixed ordering difference is visible.  
In the many-body calculation, the same $B$ and $C$ give $\mathcal V_{0.05}(20)=0.032098$ and $\mathcal V_{3.20}(20)=0.792679$.  
The operations, their common normalization, and the observation window are identical, only the input state changes.  
The one-qubit solution proves the same statement exactly, including a state in which the ordering difference is completely invisible.

The initial quadratic loss does not determine the finite-time result.  
The two endpoint states~($\lambda=0.05$ and $\lambda=3.20$) have the same $\operatorname{Var}_\psi(K)=16$ and therefore the same quadratic term, but their first numerically resolved difference occurs at fourth order and grows into a large finite-time visibility contrast.  
The leading term is fixed by the fluctuations of $K=2i[B,C]$, whereas higher ordered moments determine the later separation.

The intermediate states~($0.05<\lambda<3.20$) connect this contrast to the change of phase in the finite lattice.  
At $U=20$, the visibility remains small from $\lambda=0.05$ to $0.20$, rises sharply around the finite-size transition region near $\lambda=0.40$, and becomes large on the large-$\lambda$ side.
Thus one side of the state family nearly hides the ordering difference, while the other side reveals the same fixed difference.

The result is not restricted to one finely chosen operator.  
Other nine inequivalent pairs selected by the FunSearch-inspired search have lower visibility in $|\psi(0.05)\rangle$ than in $|\psi(3.20)\rangle$.  
For the main pair, the contrast also remains positive in all $56$ tested coefficient changes, including both one-at-a-time changes and simultaneous random changes.

This work establishes state-dependent operational invisibility of non-Abelian ordering information, which we call a form of dynamical Abelianization for the tested processes within a fixed operator pair, a specified state family, and a finite time window.  
Whether the dynamical Abelianization identified here is related to Abelian dominance remains unknown and is left for future work.
Operational visibility therefore characterizes whether ordering information is physically accessible in a specified state and process, without changing the non-commuting operator algebra that underlies it.

\section*{Data and Code Availability Statement}
The data and code supporting this work are available at \url{https://www.modelscope.cn/datasets/nbalexis/State-Dependent_Visibility_of_Non-Commutative_Ordering_in_Quantum_Dynamics}.

\begin{acknowledgments}
This work was supported in part by the National Natural Science Foundation of China under Grants Nos.~12575106 and 12147214 and by the Basic Research Projects of Universities in Liaoning Province under Grant No.~LJ212510165024.
\end{acknowledgments}

\bibliography{abelian}

\end{document}